\documentclass[12pt,a4paper]{article}

\usepackage{graphicx}
\def\ltsima{$\; \buildrel < \over \sim \;$}
\def\simlt{\lower.5ex\hbox{\ltsima}}
\def\gtsima{$\; \buildrel > \over \sim \;$}
\def\simgt{\lower.5ex\hbox{\gtsima}}
\def\gtrsim{\mathrel{\hbox{\rlap{\hbox{\lower4pt\hbox{$\sim$}}}\hbox{$>$}}}}
\def\la{\mathrel{\hbox{\llap{\hbox{\lower4pt\hbox{$\sim$}}}\hbox{$>$}}}}
\def\arcsec{\hbox{$^{\prime\prime}$}}
\def\sun{\hbox{$\odot$}}
\let\ga=\gtrsim
\begin{document}

\title{Spectroscopic confirmation of a galaxy at redshift z=8.6}

\author{M. D. Lehnert$^1$, N. P. H. Nesvadba$^2$, J.-G. Cuby$^3$, 
A. M. Swinbank$^4$, \\ S. Morris$^5$, B. Cl\'ement$^3$, C. J. Evans$^6$,
M. N. Bremer$^7$, S. Basa$^3$}

\date{}

\maketitle

\noindent
$^1$GEPI, Observatoire de Paris, CNRS, Universit\'e
Paris Diderot, 5 place Jules Janssen, 92190 Meudon, France\\
$^2$Institut d'Astrophysique Spatiale, UMR 8617, CNRS, Universit\'e Paris-Sud,
B\^atiment 121, F-91405 Orsay Cedex, France\\
$^3$Laboratoire d'Astrophysique de Marseille,
OAMP, Université Aix-Marseille \& CNRS
38 rue Fr\'{e}d\'{e}ric Joliot Curie, 13388 Marseille Cedex 13, France \\
$^4$Institute for Computational Cosmology, Department of Physics, Durham University, South Road, Durham DH1 3LE\\
$^5$Department of Physics, University of Durham, South Road, Durham DH1 3AJ\\
$^6$UK Astronomy Technology Centre, Royal Observatory, Blackford Hill, Edinburgh EH9 3HJ\\
$^7$Department of Physics, H H Wills Physics Laboratory, Tyndall Avenue, Bristol, BS8 1TL, UK\\

{\bf\noindent
Galaxies had their most significant impact on the Universe when they
assembled their first generations of stars. Energetic photons emitted
by young, massive stars in primeval galaxies ionized the intergalactic
medium surrounding their host galaxies, cleared sight-lines along which
the light of the young galaxies could escape, and fundamentally altered
the physical state of the intergalactic gas in the Universe continuously
until the present
day$^{1,2}$. Observations of the Cosmic Microwave Background$^3$, and of
galaxies and quasars at the highest redshifts$^4$, suggest that the Universe
was reionised through a complex process that was completed about a
billion years after the Big Bang, by redshift z$\approx$6. Detecting ionizing
Ly-alpha photons from increasingly distant galaxies places important
constraints on the timing, location and nature of the sources responsible
for reionisation. Here we report the detection of Ly$\alpha$ photons emitted
less than 600 million years after the Big Bang.
UDFy-38135539$^5$ is at a
redshift z=8.5549$\pm$0.0002, which is greater than those of the previously known
most distant objects, at z=8.2$^{6,7}$ and z=6.96$^8$. We find that this
single source is unlikely to provide enough photons to ionize the volume
necessary for the emission line to escape, requiring a significant contribution
from other, probably fainter galaxies nearby$^9$.}

UDFy-38135539 was selected as a candidate z$\approx$8.6 galaxy from
deep Wide Field Camera 3 observations of the Hubble Ultra Deep
field$^5$.  Its red Y$_{105}$-J$_{125}$ colour is one of the reddest
in the parent sample of $z$=8--9 candidates, and, together with the
sensitive upper limits in the optical through the Y$_{105}$ band, make
it the most plausible z$\approx$8.6 galaxy$^{5,10}$.  To search for its
Ly$\alpha$ emission, we obtained sensitive near-infrared integral-field
spectroscopic observations of UDFy-38135539 using the SINFONI spectrograph
at the ESO
Very Large Telescope, with an integration time on the source of
14.8 h in the near-infrared J-band (1.1-1.4$\mu$m).

\begin{center}
\includegraphics[width=16cm]{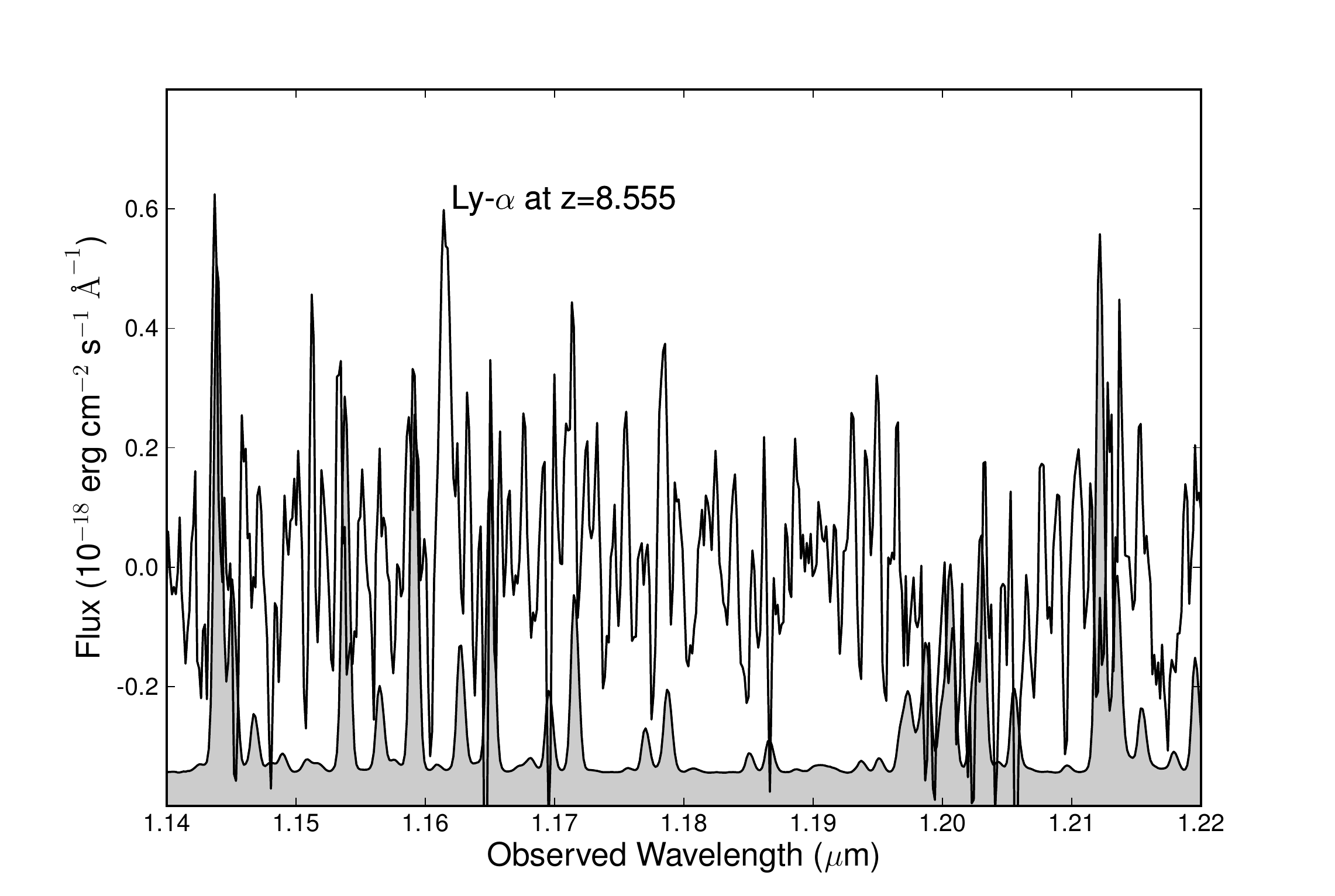}
\includegraphics[width=16cm]{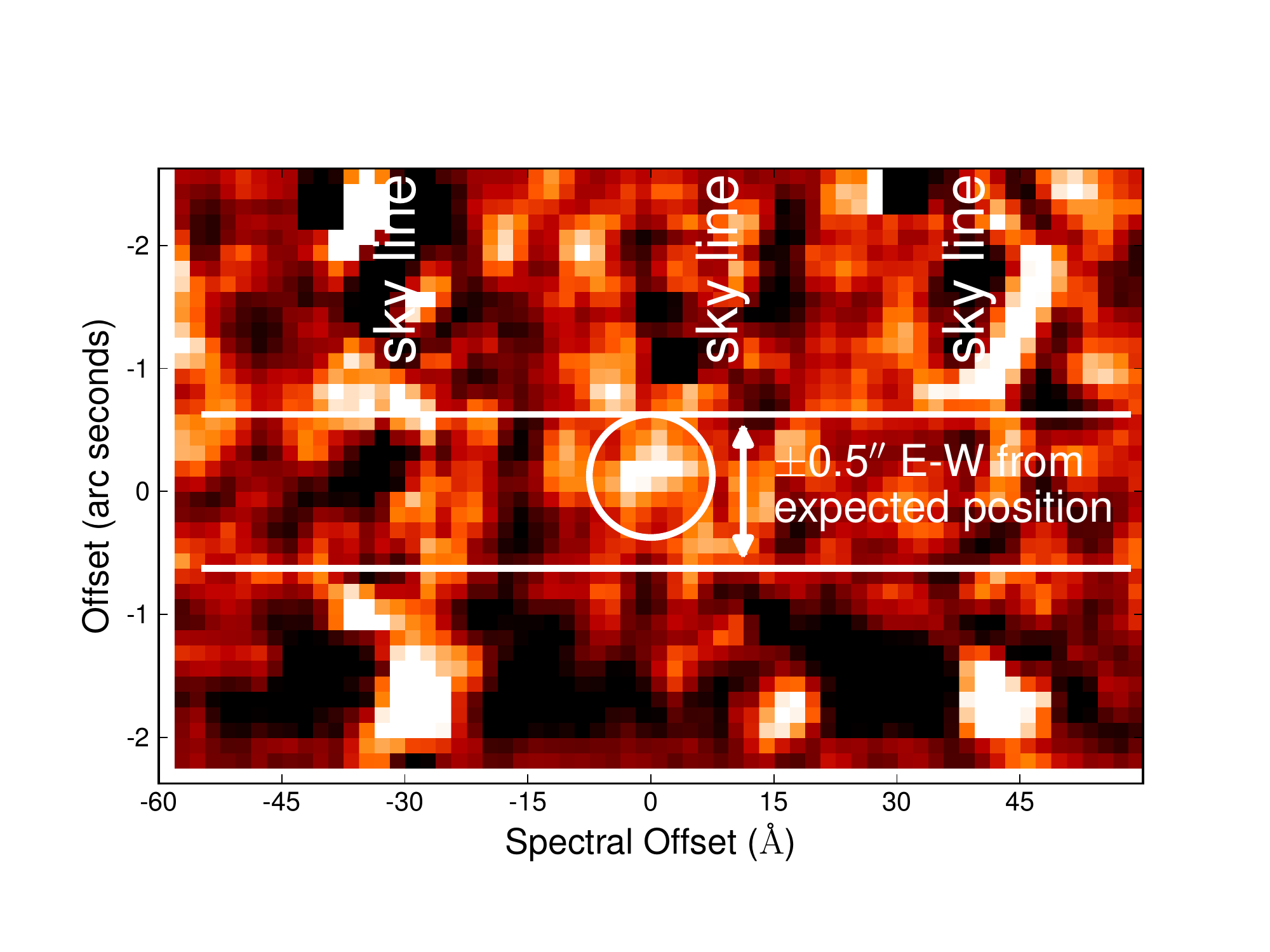}
\end{center}
{\footnotesize {\bf Figure 1 Two representations of the spectrum of UDFy-38135539
showing its significance.} {\bf a,} The spectrum shows a faint emission line
detected at 6$\sigma$ significance at a wavelength of 11,615.6\AA,
corresponding to a redshift z=8.5549$\pm$0.0020 for Ly$\alpha$. The
integrated spectrum was extracted from a square aperture of 5$\times$5
pixels, corresponding to 0.625\arcsec$\times$0.625\arcsec, which is
approximately the size of the seeing disk. The measured line full width at
half maximum is 9.2$\pm$1.2\AA, which is about 1$\sigma$ greater than the
instrumental resolution. The line flux is (6.1$\pm$1.0)$\times$10$^{-18}$
erg cm$^{-2}$ s$^{-1}$, detected at 6$\sigma$ significance.  All of the
line parameters (redshift, width, flux and significance) were estimated with
a Monte Carlo simulation assuming a Gaussian line and randomly generated
Gaussian noise similar to that estimated for the observed spectrum. We
note that the absolute flux calibration may have a significant systematic
uncertainty of up to 30-40\%, but this does not affect the estimate of
the significance of the line detection. The night sky spectrum, scaled
arbitrarily, is shown in grey.  Regions of particularly deviant values
in the spectrum correspond to strong night sky lines.  The emission
line from the source lies fortuitously in a region relatively free of
night sky contamination.  We estimate that the percentage of regions
in the night sky with a background as low as that near the detected
line is approximately 50\% for 1.15-1.35$\mu$m and is generally lower
over the rest of the SINFONI J bandpass.  {\bf b,} The sky-subtracted
two-dimensional spectrum shows the projection of the spectrum along the
spectral and right-ascension axes of the data cube.  It corresponds
to a two-dimensional longslit spectrum obtained with a slit width of
0.625{\arcsec} positioned along right ascension on the sky.  The object
is indicated by the white circle, the regions affected by the night sky
lines are labelled, and the range in the expected position of the source
is marked.}

In our spectrum, we detect faint line emission at a wavelength
of $\lambda$=11615.6$\pm$2.4\AA, corresponding to a redshift
z=8.5549$\pm$0.0002 assuming that the line is Ly$\alpha$ (Fig.~1).
This redshift is consistent with the redshift estimates made by comparing
the photometry to models of spectral energy distributions$^{10}$.
We constructed a line image by spectrally summing a data volume containing
the line (Fig.~2).  Both the size of the emission and spectral width of
the line are what would be expected for a source of astrophysical origin.
If the line were due to detector noise, this would generally lead to
a line-width and source size that are smaller than the resolution
of the spectrograph and the smearing due to atmospheric turbulence
(Supplementary Information).

The photometry from the Hubble Space Telescope (HST) allows for an
alternative (but unlikely) redshift of z=2.12, so we also investigate
whether the emission line could be another emission line at lower
redshift.  In this case, the line may be the [O{\sc ii}] $\lambda$3,726
and 3729 emission doublet, but we rule this out because the [O{\sc
ii}] doublet would be clearly resolved, and, hence, the line would be
intrinsically wider than observed (details are given in the Supplementary
Information).

\includegraphics[width=12cm]{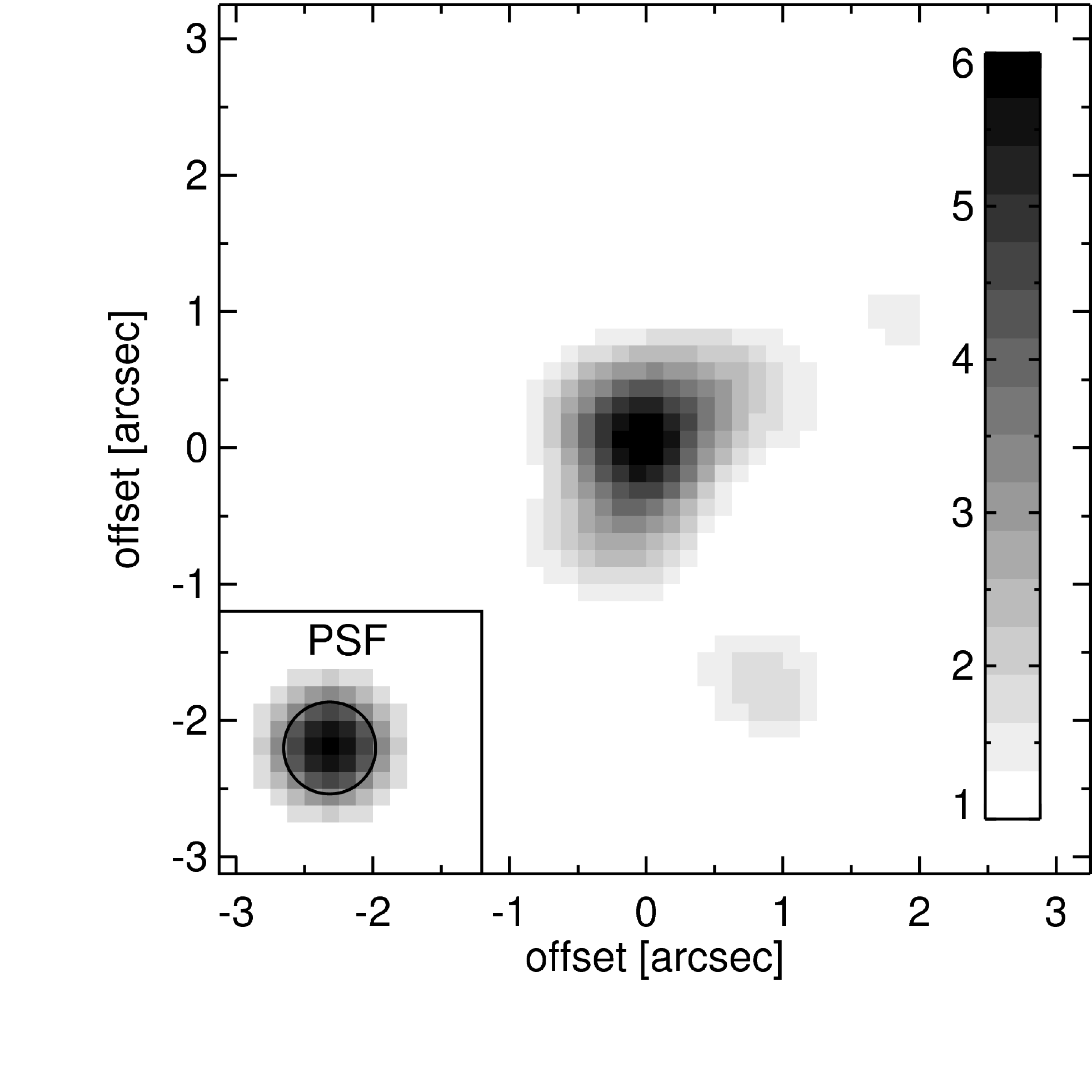}\\
{\footnotesize {\bf Figure 2 Lyman-$\alpha$ line image of UDFy-38135539.} The line image
was constructed by summing the region containing the emission in the
wavelength direction.  The inset shows the expected morphology of a point
source with the same signal-to-noise ratio in its centre as that of the
source and the circle shows the size of the point spread function (PSF).
The colour bar shows the significant relative to the root mean squared
noise in the data set.  The image has been smoothed by a Gaussian with the
same width as the point spread function.  The size of the line image is
consistent with the expected size of an intrinsically unresolved source
whose image is smeared out by the turbulence in the Earth's atmosphere
and distortions induced by the telescope and instrument optics.}

For a standard $\Lambda$ cold dark matter ($\Lambda$CDM) cosmology, the
Hubble constant is 70 km s$^{-1}$
Mpc$^{-1}$, the dark matter density is 0.3 and the dark energy density
(cosmological constant) is 0.7,
the luminosity distance is d$_l$=86.9 Gpc, and the total
flux of Ly$\alpha$ emission implies that the luminosity of the source
of 5.5$\pm$1.0$\pm$1.8$\times$10$^{42}$\,erg\,s$^{-1}$
(1-sigma uncertainty and systematic uncertainty).
In comparison, currently known Ly$\alpha$ emitters over
a wide range of redshifts (z$\approx$3-7) have typical luminosities of
(3-10) $\times$  10$^{42}$\,erg\,s$^{-1}$ without
significant evolution$^{11,12}$.  Thus UDFy-38135539 can be considered
a typical Ly$\alpha$ emitting galaxy.

At z=8.55, the observed H$_{160}$-band window samples around 1,700\AA\
in the rest-frame ultraviolet.  The H-band magnitude implies a flux
density of log\,(f$_{1700\AA}$ (erg\,s$^{-1}$\,cm$^{-2}$\,Hz$^{-1}$)) =
$-$30.7$\pm$0.2 and an intrinsic luminosity density of log (L$_{1700\AA}$
(erg s$^{-1}$ Hz$^{-1}$))=28.3$\pm$0.2.  The continuum luminosity density
is about one magnitude fainter than M$^{\star}_{UV}$, the characteristic magnitude of the ultraviolet luminosity function, for galaxies with
redshifts $z$=6-7$^{13,14}$.  If the observed evolution of the luminosity
function from z$\approx$3 to z$\approx$7 continues to z$\approx$8.6, this would
imply that UDFy-38135539 is a typical M$^{\star}_{UV}$ galaxy$^5$.

Although we can not derive the characteristics of the stellar population
in UDFy-38135539, similarly-selected galaxies at lower redshifts,
z$\approx$5-7, seem to be young (ages between 10 and 100 million years,
Myrs) and have both low metallicity and low extinction$^{15--17}$.
Plausible characteristics for the stellar population in UDFy-38135539
include low metallicity (10\% of the solar metal abundance to essentially
zero heavy elements), a mass distribution either similar to that of
massive stars in local galaxies or with only very massive stars$^{18}$,
and ages between 10 and 100 Myrs (but perhaps as old as 300 Myrs$^{19}$).
Considering this range of mass distributions, metallicities and ages,
we estimate the star-formation rate of UDFy-38135539, on the basis of
the UV continuum luminosity, to be 2-4 M$_{\sun}$ yr$^{-1}$.  On the
basis of the Ly$\alpha$ luminosity, we estimate the star-formation rate
to be 0.3-2.1 M$_{\sun}$ yr$^{-1}$.  However, we caution that, owing
to the unknown absorption by the intergalactic or interstellar medium,
the star-formation rate estimated using this Lyman-$\alpha$ luminosity
should be considered as a lower limit.

Observing Ly$\alpha$ emission in a galaxy at z=8.55 suggests that the
surrounding inter-galactic medium must be ionized beyond $\sim$1\,Mpc from
the source to allow the emission to escape.  We note that the
mean recombination time at z$\sim$8.6 is approximately a Hubble time at
this redshift ($\sim$600\,Myr).  Thus, once a region of the inter-galactic
medium becomes ionized, we expect that it will be fossilised because the
gas has insufficient time to recombine before the end of reionization.
Moreover, because the time during which a source is a luminous emitter
of ionizing photons is significantly less than a Hubble time, the sources
that created any ionized bubble in the intergalactic medium may be very
difficult to detect$^{20}$.

It is therefore instructive to investigate the possible size of the
ionized region around UDFy-38135539.  The total number of ionizing photons
that UDFy-38135539 has produced allows us to estimate the size of the
bubble it has ionized$^{9}$.  Adopting the star-formation rate derived
from the ultraviolet continuum flux density, and assuming the range
of characteristics as discussed above, we estimate that UDFy-38135539
will ionize a region of between $\sim$0.1(f$_{esc}$/0.1)$^{1/3}$
and 0.5(f$_{esc}$/0.1)$^{1/3}$ Mpc in radius where f$_{esc}$ is the
escape fraction of ionizing photons.  With such a small radius, the
neutral intergalactic medium surrounding the bubble will significantly
suppress the intrinsic Ly$\alpha$ emission from the source$^{21,22}$
(Fig.~3).  Outflows of gas driven by the star-formation within the
source may help the escape of Ly$\alpha$ radiation$^{23}$.  However,
for the line emission to be highly redshifted relative to the systemic
velocity of the sources, the optical depth to Ly$\alpha$ must be high,
further suppressing the emission.

Given the difficulty of ionizing the surrounding medium sufficiently,
the most likely explanation for the relatively strong Ly$\alpha$
emission from UDFy-38135539 is that other sources within a few megaparsecs
of this source may also have contributed to ionizing this volume.
Indeed, the likelihood that many sources contribute to the ionization of
bubbles during reionisation has been discussed previously$^{20,24}$.
As the intergalactic medium is potentially 30-50\% ionized at this
redshift$^{25}$, it would not be surprising that multiple ionizing sources
exist in the vicinity of UDFy-38135539 that may also be responsible for
ionizing a significant fraction of the entire volume of the intergalactic
medium.

Thus, the luminous Ly$\alpha$ emission of UDFy-38135539 implies that it is
probably surrounded by other sources that contributed to the reionization
of the Universe, but which have not yet been discovered or characterized.
With the current capabilities of the Wide Field Camera 3 and sensitive
spectrographs in the near-infrared on 8-10-m-class telescopes such as
SINFONI (and, in the near future, the KMOS spectrometer, also at the Very
Large Telescope), it may be possible directly to detect and characterize
the galaxies during the epoch of reionization, helping us to understand
how galaxies reionized the Universe. However, the biggest breakthroughs
in our understanding of galaxies responsible for reionization will come
with observations using the European Extremely Large Telescope and the
James Webb Space Telescope over the next decade.

\begin{center}
\includegraphics[width=12cm]{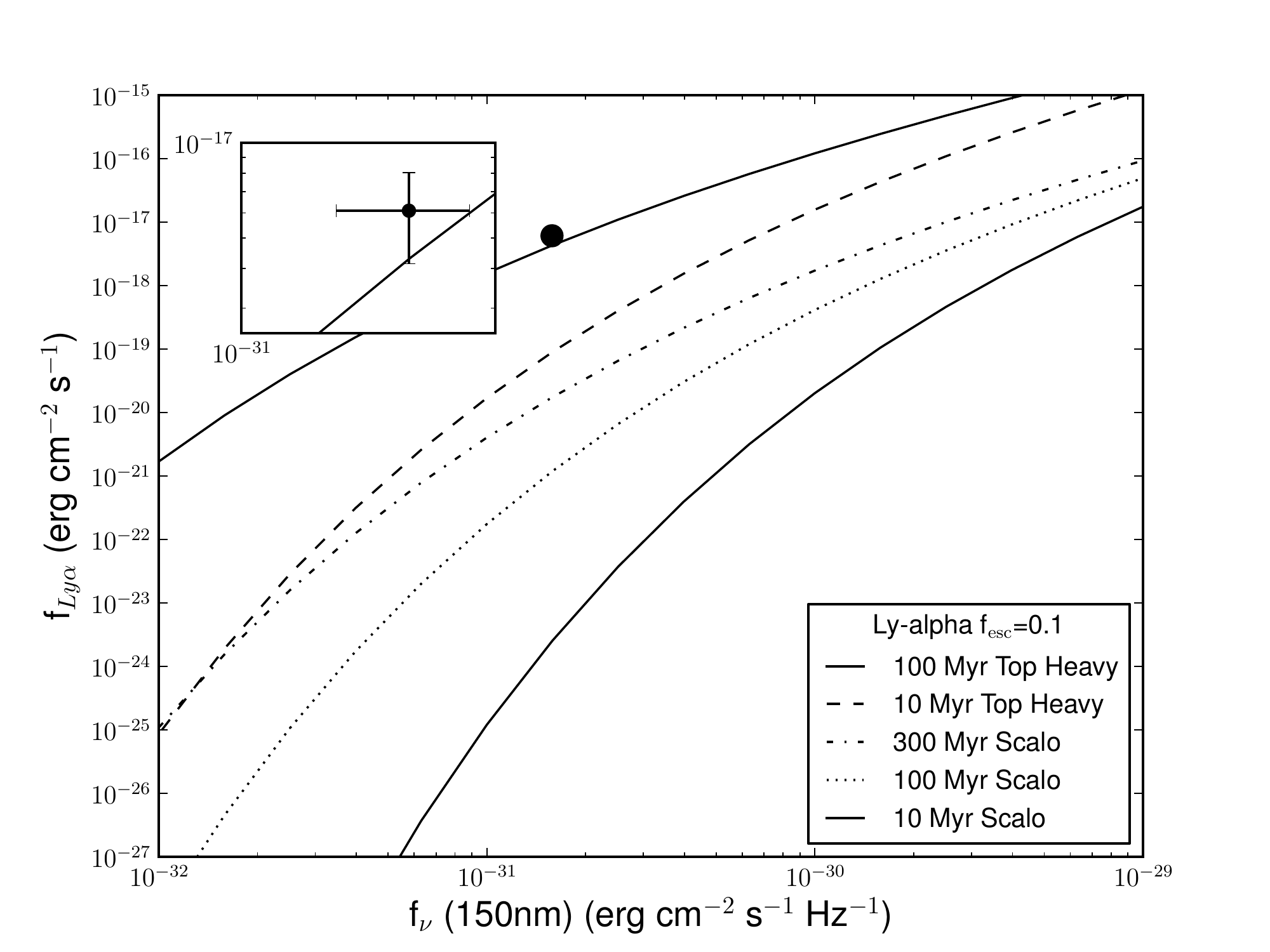}
\end{center}
{\footnotesize {\bf Figure 3 The predicted Ly$\alpha$ flux for a given
ultraviolet flux density.} The Ly$\alpha$ flux, f$_{Ly\alpha}$, is
predicted assuming a range of characteristics for the stellar population
within UDFy-38135539.  The characteristics -- age, metallicity and
distribution of stellar masses -- determine the relationship between
the non-ionizing ultaviolet continuum at 1500\AA\ (with flux density
f$_{\nu}$) and the ionizing continuum with $\lambda$ $<$912{\AA}. We
adopted a range of ages from 10 to 300 Myr for a metal poor stellar
population given by a Scalo initial mass function$^{26}$. Our other
initial mass function is the one that only contains massive stars,
$>$ 100 M$_{\sun}$, which have zero metallicity$^{18}$. For this top
heavy initial mass function, we only considered ages of 10 and 100 Myrs
because it is unrealistic for metal-free star-formation to persist for
after the first supernova explosions which are expected after a few to
a few tens of megayears.  For all the calculations, we have assumed an
escape fraction of ionising photons of 10\%.  Estimates of the escape
fraction in the local Universe up to about z$\sim$3.3 suggest modest
fractions of $\sim$10\% or less$^{27-29}$.  The black circle represents
the UV continuum flux density and Ly$\alpha$ flux of UDFy-38135539.
The uncertainties in this data are shown in the inset (the 1-$\sigma$
random uncertainty and the systematic uncertainty added in quadrature).
The uncertainty in the Ly$\alpha$ flux is dominated by the systematic
uncertainties which are included in the error bars in the inset.  Under
our assumptions, the Ly$\alpha$ flux of UDFy-38135539 is greater than
that expected if it alone was responsible from ionising its local volume.
Because the recombination time is long at z$\approx$8.6, $\sim$600 Myrs,
many of the sources responsible for ionising this region could have
easily faded or may simply be of lower luminosity.}

\pagebreak

%\bibliographystyle{nature}
%\bibliography{J1808burst}

\smallskip
\noindent {\small {\bf Supplementary Information} is linked to the online
version of the paper at www.nature.com/nature.

\smallskip
\noindent {\small {\bf Acknowledgements} We thank the Director General
of ESO for generous allocation of time and the staff on Paranal
for conducting the observations.  We also thank F. Combes,
S. Zaroubi, M. Haehnelt, D. Valls-Gabaud, and J. Dunlop for
discussions.}

\smallskip
\noindent {\small {\bf Author Contributions} M.D.L. led the writing
of the paper and the presentation of the results and was responsible
for the modeling shown in Fig. 3. N.P.H.N designed the observations,
reduced all of the data and was responsible for the data shown Figs
1 and 2.  A.M.S., J.-G.C., B.C., and S.B. also examined the data. S.M.,
M.N.B., N.P.H.N., A.M.S. helped significantly editing the manuscript. All
authors discussed the results and commented on the manuscript.

\medskip
\noindent {\small {\bf Author Information} Reprints and permissions
information is available at www.nature.com/reprints. The
authors declare that they have no competing financial interests.
Correspondence and requests for materials should be addressed
to MDL (e-mail: matthew.lehnert@obspm.fr) or N.P.H.N. (email:
nicole.nesvadba@ias.u-psud.fr).}

\pagebreak

\def\ltsima{$\; \buildrel < \over \sim \;$}
\def\simlt{\lower.5ex\hbox{\ltsima}}
\def\gtsima{$\; \buildrel > \over \sim \;$}
\def\simgt{\lower.5ex\hbox{\gtsima}}
\def\gtrsim{\mathrel{\hbox{\rlap{\hbox{\lower4pt\hbox{$\sim$}}}\hbox{$>$}}}}
\def\la{\mathrel{\hbox{\llap{\hbox{\lower4pt\hbox{$\sim$}}}\hbox{$>$}}}}
\def\arcsec{\hbox{$^{\prime\prime}$}}
\def\arcmin{\hbox{$^{\prime}$}}
\def\sun{\hbox{$\odot$}}
\let\ga=\gtrsim

\noindent
{\bf\huge Supplementary Information}

\bigskip
The emission line presented here could have several plausible origins
other than Lyman-$\alpha$, including non-random detector noise, night-sky
residuals, or another astrophysical emission line at lower redshift.
Here we discuss the reduction methods, and provide several tests of the
detection, to demonstrate why it is Lyman-alpha emission.

\section*{S1 Observations and Data Reduction}

We used SINFONI$^{31}$ on UT4 of the VLT to obtain deep, 3-dimensional
spectroscopy of UDFy-38135539.  We observed UDFy-38135539 in the J
band, which covers the expected wavelength of Ly$\alpha$ at a redshift
z$\sim$8.5, the redshift determined from the multi-wavelength photometry
available for this source$^{32}$.  We observed for 16~hrs, granted as
Director's Discretionary Time (program ID 283.A-5058), and obtained a
total of 53400 seconds of integration time on the source. Individual
exposure times were 600s, and we adopted a dither pattern where the
source remained in the field of view for all exposures. Observing
faint, high-redshift galaxies requires a high accuracy in the pointing,
better than 1\arcsec. We therefore defined our pointing by acquiring
a star at about 1.5\arcmin\ from the position of UDFy-38135539$^{32}$
at the beginning of each Observing Block (a sequence of 5-6 individual
exposures with a total length of 1 hr). At the end of each Observing
Block we observed a nearby star of known J-band magnitude and spectral
type at a similar airmass. Flux scales were obtained from these stars,
and we also used the size of the point spread function of these stars
to monitor the seeing. 

An additional uncertainty can arise in the calibration due to both the
variation in the atmospheric absorption at the wavelength of the line
and to the crudeness of the flux calibration.  The depth of absorption
in the spectral region near the line is significant, $\approx$10-30\%,
and is variable depending on the amount of water vapour in the atmosphere
and the telescope elevation during which the observations were taken. The
data and standard stars are taken at somewhat different times and air
masses, adding additional uncertainty.  Spectrophotometric calibration in
the near-infrared is not very accurate. There are no generally available
spectrophotometric standards as in the optical and the calibration star
is used to simultaneously provide flux calibration and to remove the
effects of atmospheric absorption. The crudeness of the calibration
probably induces an additional uncertainty of 20-30\% and changes in
the atmospheric absorption between the observations of the source and
calibration star can introduce systematic uncertaintes of about the same
order.  This additional uncertainty does not influence the significance
of the detection, as the measurement of the noise and the line flux are
affected by the same systematic uncertainties, but it does influence the
accuracy of the estimate of the line flux.  Systematic uncertainties in
the absolute flux measurement are therefore of-order 30-40\%.

The instrumental resolution was measured from the widths
of night sky lines extracted from a SINFONI cube near the wavelength of
the line, and the error estimates are based on a Monte-Carlo simulation
of the data with a Gaussian line profile and the noise characteristics
of the region near the line. To determine the instrumental
resolution, we used a data cube obtained in exactly the same manner as
the science data with the only difference being that we did not subtract
the night sky.  The spectral resolution was measured to be 190 km s$^{-1}$
or R=$\lambda$/$\Delta\lambda$= 1580 at $\lambda\sim$1.16$\mu$m.

The data were reduced using our own data reduction software based on
IRAF$^{33}$.  These routines have been tested and used extensively and
have been discussed extensively in previous articles$^{34,35,36,37}$.
A study similar to this one has been attempted previously$^{38}$.

\section*{S2 The Nature of the Line}

An obvious concern in searching for faint line emission from galaxies
at the highest redshifts is that the detection is spurious.  This is
particularly worrisome for near infrared observations. Near-infrared
detectors have pixels that do not behave linearly with exposure and/or
have a significantly different response from the vast majority of pixels.
In addition, the line and continuum emission from the sky emission has
structure and is also variable.  These effects result in reduced data
that do not have strictly Gaussian random noise.  These effects can
be largely, but not entirely corrected for during the data reduction.
Because of this, it is important to make a number of comparisons between
the properties of any putative emission line and those expected for
a real signal.

To show that the line is not an instrumental artefact, we have made
several tests that an astrophysical line must meet at a minimum to be
considered real.

\subsection*{S2.1 Characteristics of the Line}

The line in the integrated spectrum has a width that is slightly larger
than the instrumental resolution.  The characteristics of the line
detected in our cube are given in Table 1. While we would expect the
Ly$\alpha$ profile to be intrinsically asymmetric due to the absorption
on the blue side of the profile due to neutral Hydrogen, the observed
line is not resolved and thus has a Gaussian profile.  Observations of
Lyman-$\alpha$ emitting galaxies at redshifts around 5 also show symmetric
profiles when they are approximately unresolved, even with much higher
signal-to-noise than 6$^{39}$. The faintness of the signal and low
signal-to-noise makes it impossible to detect the faint, and possibly
non-Gaussian wings often observed in Ly$\alpha$ emitting galaxies.

The extent of the line image is 0.6\arcsec$\times$0.7\arcsec, similar to
the size of the seeing disk which results from atmospheric turbulence. The
WFC3 image$^{32}$ shows that the source has an intrinsic size in the
continuum of 0.3\arcsec. If the emission line region of the source is
not larger than the continuum source by about a factor of 2, it would
be unresolved in our data as observed.

The line is found in a region of the night sky that is relatively free
of bright night-sky lines.  Night-sky line residuals, due to mechanical
instabilities of the instrument leading to small (fractions of a pixel)
wavelengths shifts, can lead to features that may be mistaken for real
emission lines.  The brightest of these spurious features arise in a part
of the data cube that is near the edge of the slitlets of the slicer.
In Fig.~1 of the main article, the night sky line residuals are more
prominent near the periphery of the spectrum. The candidate Lyman alpha
emitter is in a region of the cube that is unaffected by the edges of
the slitlets.

The line emission is within 0.4\arcsec\ of the expected position of the
source in our data cube, and within about one half of the size of the
point spread function.  Due to the low signal-to-noise of the detection
itself, uncertainties in the relative astrometric position of the offset
star and target, and inaccuracies in the telescope offset, imply that
the offset we observe in our cube is consistent with the position of
the source.

\begin{table}
\begin{tabular*}{1.0\textwidth}{@{\extracolsep{\fill}} lcccc}
\hline
ID&$\lambda_{obs}$ & redshift & FWHM      &  Flux \\
\hline
Ly$\alpha$& 11615.6$\pm$2.4 & 8.5549$\pm$0.0002 & 9.2$\pm$1.2 & 6.1$\pm$1.0\\
\hline
\end{tabular*}
\caption{Emission-line properties measured from our SINFONI data
of UDFy-38135539. Column (1) -- Line ID.  Column (2) -- observed
wavelength. Column (3) -- Redshift assuming the line is Ly$\alpha$ at
1215.67\AA. Column (4) -- Measured full width at half maximum (FWHM) in
units of \AA. The line is unresolved.  Column (5) -- Line flux in units
of 10$^{-18}$ erg cm$^{-2}$ s$^{-1}$. The uncertainty in the line flux
is derived from repeated realisation of the line given the noise in the
data set. There is a systematic uncertainty of the flux of 30-40\% which
is not included in the estimate given.} \label{tab:sinfobs} \end{table}

\subsection*{S2.2 Random or Correlated Noise?}

All of these suggest at a minimum that the emission line is consistent
with being of astrophysical origin from the targeted source.
Even though it appears that this is likely to be an emission line
from a source, it could also be that this is a rare noise spike in the
data that coincidently happens to have characteristics expected for an
astrophysical source.  We will now illustrate that the line with the
properties we have estimated is also rare, rare enough not to be associable
with random or correlated detector noise.

In our three-dimensional data cube we estimate that there are
approximately 500,000 pixels where we expect night-sky line residuals
that are similar to those in the area covered by the source (and which
are contained in all exposures). We used a simple Monte Carlo approach
to investigate how many of these pixels could produce a spurious noise
signal similar to the actual signal that we have observed.

We randomly selected a spatial position covered by the data cube and
extracted the 25 one-dimensional spectra centred on that position (in a
5 by 5 spatial pixel region, specifically chosen to be approximately the
size of one seeing disk). We then randomly selected a wavelength within
the region covered by those spectra and, for each spectrum, fitted a
Gaussian line profile centred at that wavelength. The width of the
fitted Gaussian was constrained to be between the spectral resolution
and 400 km s$^{-1}$ and its total flux was allowed to vary between 3 and
90 times the noise level. We then determined how many of the 25 spectra
had acceptable fits comparable to the detected feature.

This procedure was repeated 25000 times and on 301 occasions (1.2\%)
resulted in reasonable fits for one or more of the 25 extracted spectra
(see Fig.~\ref{fig:multiplicity}). The maximum number of spectra
with acceptable fits in any one trial was eleven, found in a single
trial. There was not a single instance of all 25 spectra resulting
in acceptable fits, as there was for the detected feature.  Given the
statistics shown in Figure 1, the probability of generating such a feature
purely from the noise characteristics of the data cube is at best one in
25000 (a confidence level of $>99.99\%$), and likely considerably lower.

\begin{figure}[t]
\centerline{\includegraphics[width=5in]{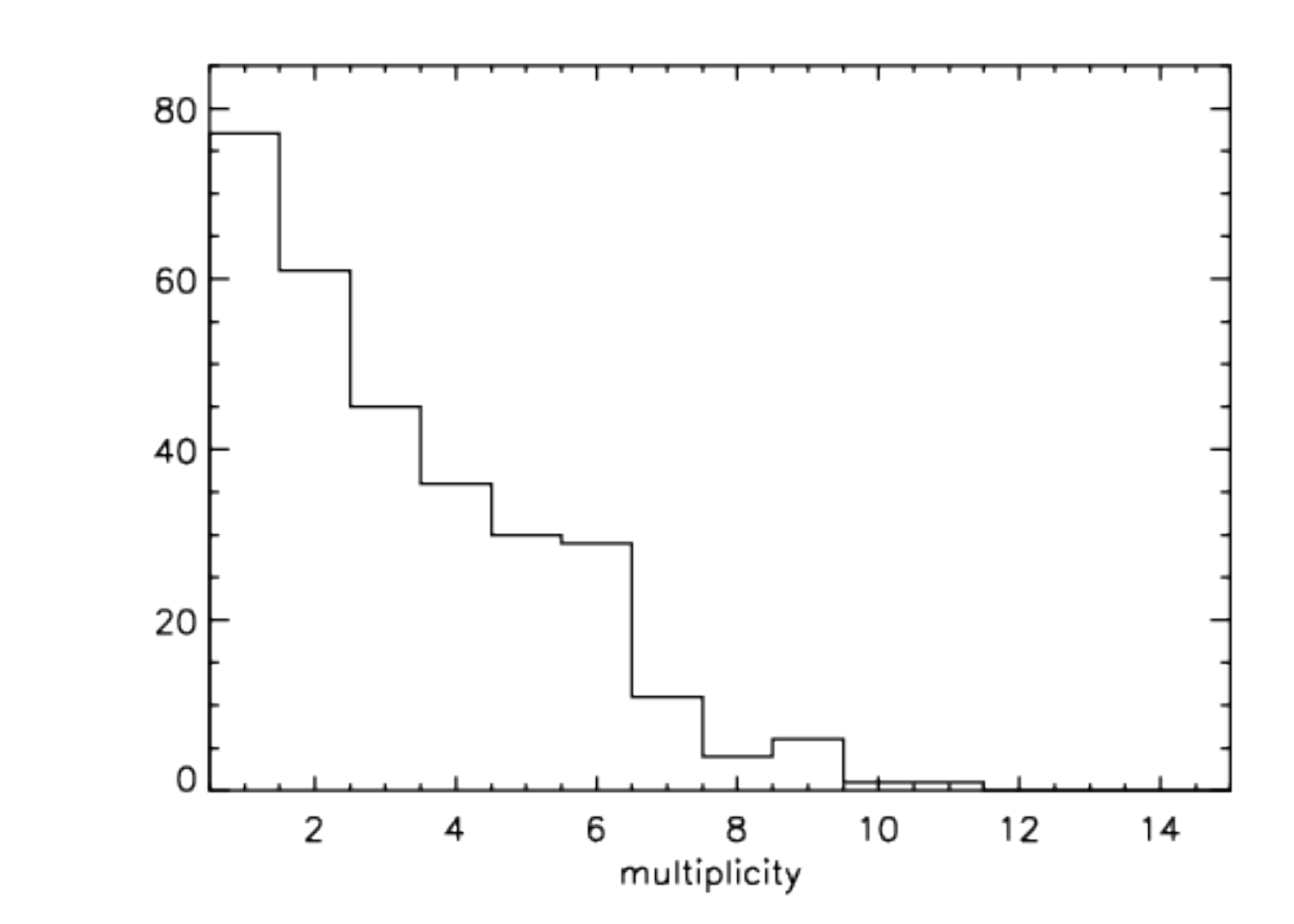}}
\caption{The ``multiplicity'' distribution.  The multiplicity is the
number of pixels that can adequately be fit with similar fit parameters
within a box corresponding to the size of the seeing disk (5$\times$5
pixels), surrounding a randomly chosen pixel in a region containing
no signal from and have noise characteristics similar to that of the
region around the detected line emitter.  The detected line emitter
has a multiplicity of 25, whereas in the noise has multiplicities
$\le$11. Having found a multiplicity of 11 in 1 trial of 25,000, we
can rule out larger multiplicities due to noise at a confidence level
$>$99.996\%.}
\label{fig:multiplicity}
\end{figure}

\subsection*{S2.3 A Spurious Line: Negative Line Image}

There are several other ways to argue that the line is not likely to
be spurious.  We searched for the negative images of the line which are
a by-product of our observing strategy, and allow for a particularly
powerful test of the astrophysical origin of our signal. We adopted a
dither strategy where the source is within the field of view at all
times, but falls into different parts of the detector in subsequent
individual exposures. This means that we can use one object frame to
subtract the sky from another, and the presence of the source in each of
these frames implies that we produce a characteristic negative signal
in each individual sky-subtracted frame, where we subtract the source
from an empty part of the sky. We will refer to this negative signal as
the ``shadow''. Identifying the shadow is a particularly powerful test,
since the shadow arises from a different part of the detector, sharing
only half the pixels with the source in each individual exposure. From
our experience with integral-field spectroscopy with this instrument
of over a 100 faint galaxies, we are not aware of any data cube where
a shadow would have been produced by an instrumental artefact.

Since the shadow does not fall in the same region of the detector for all
dithers, and because the line is so faint, we reconstructed a dedicated
``shadow cube'' from the individual frames. We calculate the expected
position of the shadow in each individual exposure from the measured
target position. We then shift the individual cubes such that all shadows
fall on the same position and combine the individual exposures to obtain a
shadow cube, where the residual of the target falls at the same position
as the galaxy in the original combined cube (Fig.~\ref{fig:shadows}).
The shadow line is clearly seen at the wavelength of the detected line
(Gaussian line profile above the spectrum), and with a similar width
and flux.

\begin{figure}[t]
\centerline{\includegraphics[width=5in]{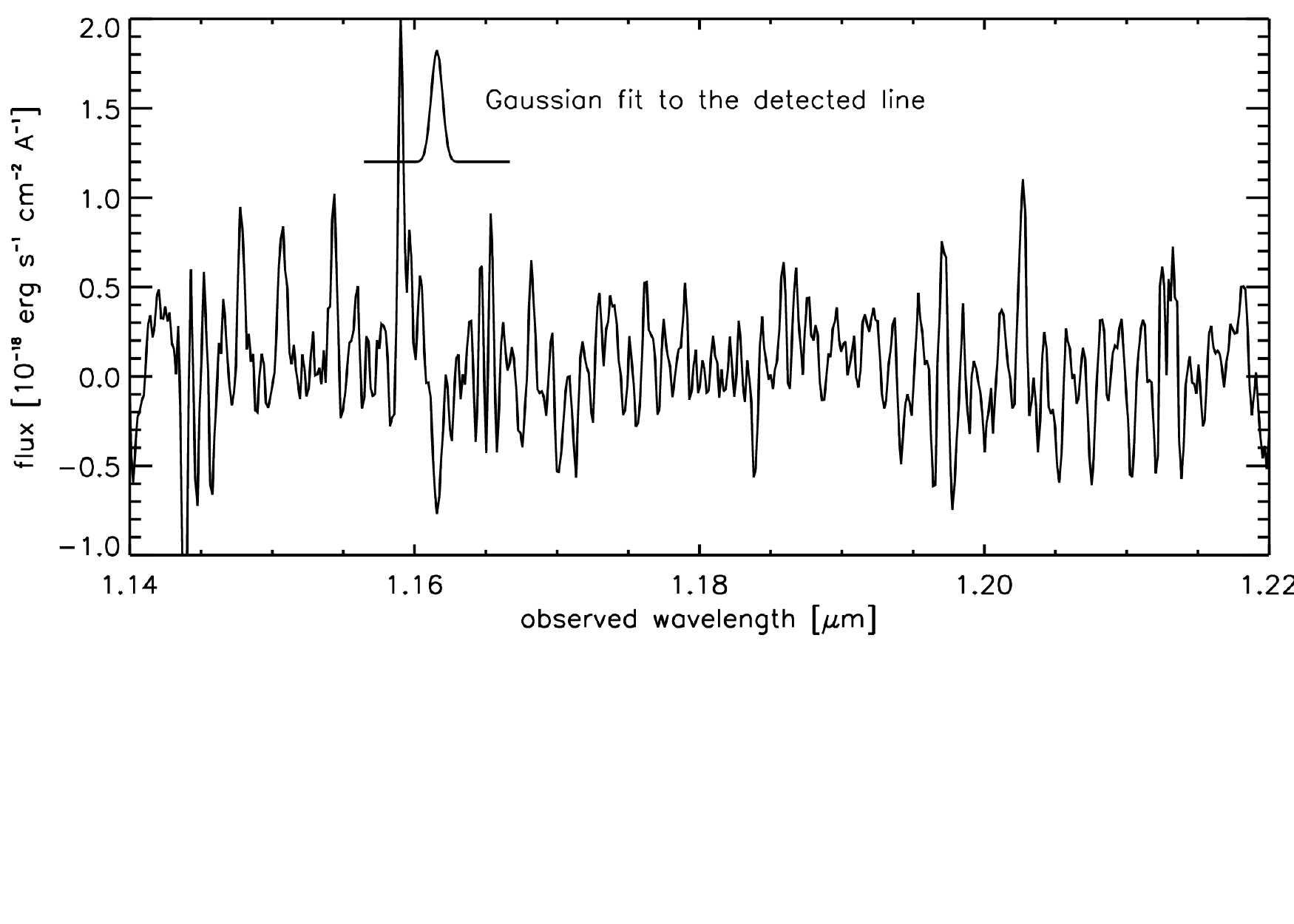}}
\caption{The integrated spectrum of the ``shadow'' in the data set. Our
strategy for subtracting the sky and combining the data set produces a
negative residual due to the presence of the source in all sky-subtracted
frames. These shadow shares all of the exposures at the position of
the source, but a completely different set of pixels in the ``off''
position compared to the positive image.  The Gaussian profile marks
the fit of the Lyman-$\alpha$ emission line and the negative residual
is clearly visible.}
\label{fig:shadows}
\end{figure}

We also evaluated the individual frames to infer whether the candidate
emission line may be dominated by non-Gaussian noise originating from
individual frames. We followed two approaches. We constructed two cubes
each containing half of the frames. For Gaussian noise we expect to
find a signal-to-noise ratio lower by $\sqrt 2$ in each subset, giving
SNR$\sim$4.6$\sigma$. We extracted integrated spectra at the same position
and with the same apertures as for the full data set, and measured the
signal-to-noise ratio of each line. We find signal-to-noise ratios of
4.3$\sigma$ and 3.5$\sigma$ for each subset, respectively. Given that
the faintness of the detection alone makes each estimate accurate only
within $\sim$25\% this is fully consistent with what is expected for
half of the full data set.

To further evaluate whether the signal may be dominated by strongly
deviant pixels in individual frames, we also extracted line images
for each individual, fully reduced data cube, by collapsing over the
wavelength range near the wavelength of our line detection. In total
we cover a range of 30{\AA} (3$\times$ the full width at half maximum
(FWHM) of the line) to be conservative. (By conservative, we mean
that we took a spectral range much larger than necessary to extract
the signal and therefore will possibly include additional sources of
non-random noise). We then inspected each line image individually,
in order to identify frames which were affected by bad pixels and/or
night-sky line residuals at the position of the source. We found
that 20 of the frames {\sl could} be contaminated. However, since
the line image covers a range of 30\AA, which includes a faint night
sky line residual slightly red-ward of the source, this does not imply
necessarily that the line itself is affected. We combined all 20 of the
contaminated frames (which had non-Gaussian noise near the position of
the source) and also the remaining 69 frames that were not so affected
(i.e., where the noise was better behaved and approximately Gaussian).
We found a strong, narrow (FWHM$\sim$1.3 spectral pixels) spike in the
cube constructed using the frames that were contaminated, but at the
wavelength of a night-sky line, and recovered a faint line consistent
with that found in the full data set in the combined cube of individual
frames that were not contaminated. Within the larger uncertainties, this
cube of uncontaminated frames also produced a line with a flux consistent
with that found when using the full data set. We thus conclude that the
frames with strongly non-Gaussian noise do not have a significant impact
on our line measurement in the full cube.

\section*{S3 Other Possible Sources of Line Emission}

Having shown that instrumental effects are unlikely to cause the
observed emission line, could it be a line due to a source at lower
redshift that is not Lyman-$\alpha$?  The very red colour and lack
of detection in the optical (B$_{435}$$<$29.8, V$_{ 606}$$<$30.2,
i$_{775}$$<$29.8, z$_{850}$$<$29.1 all 2-$\sigma$ limits)$^{40}$ makes
it highly unlikely that UDFy-38135539 could be an emission line galaxy
at lower redshift$^{32,40}$.  The best fit redshift for the photometry
is z=8.45 with an acceptable range of z=7.75-8.85. There is also a
shallow (reduced $\chi^2_{\nu}=3$) secondary minimum in the probability
distribution at z=1.60-2.15, which could be caused by a large spectral
break at 4000\AA\ in the rest-frame of the galaxy$^{40}$.

The much fainter and more stable backgrounds of the WFC3 on board the
{\em Hubble Space Telescope} compared to ground-based near-infrared
imagers make the photometry particularly robust. However, additional
systematic uncertainties are possible, for example line contamination
in the filter bandpass used to discover the source. Our measured line
flux, taken at face value, would lower the intrinsic J$_{125}$ magnitude
from J=28.41$\pm$0.24 to J$_{125}^{corr}$=29.7$\pm$0.3, implying a bluer
intrinsic continuum J-Y colour, Y$_{105}$-J$_{125}^{corr}$ $>$1.0$\pm$0.3
(1-$\sigma$), and a lower redshift is possible. This illustrates the
importance of emission-line measurements for candidate Lyman-break
galaxies at the highest redshifts, and makes it worthwhile to specifically
test all alternative redshifts that may be implied by the line detection.

The only other plausible line identifications at lower redshifts are
the generally bright lines from galaxies of H$\alpha$ at 656.3nm, which
would imply a redshift of z$=$0.77, [OIII] at 500.7nm, which would imply
a redshift of z$=$1.32, and the [OII] doublet at 372.6 and 372.9 nm,
which would imply a redshift of z$=$2.12. Given the secondary minimum in
the photometric redshift distribution around z$\sim$2$^{40}$, we must
make detailed tests of the possibility that the detected line could be
[OII] at z$=$2.12. For illustration purposes only, we have added an
artificial doublet line profile to our data set after subtracting off
our best fit profile from the line (Fig.~\ref{fig:o2}).  The figure
shows that any [OII] doublet emission should be both resolved and wider
than the current best fit single line.  To go beyond this illustration
and to show statistically that the [OII] doublet is inconsistent
with the properties of the line, we carried out a simple Monte-Carlo
simulation.  For this we generated 10$^6$ artificial data sets with the
noise characteristics of our data cube, and added two delta functions
separated by 2.8(1+z$_{[OII]}$)\AA\ which were then convolved with two
Gaussian profiles with the resolution of our data. We assumed a line
ratio R$_{3726/3729}=$1. In a vast majority of the cases ($\sim$99\%)
we see a resolved doublet, and in all cases, a width of the doublet that
is significantly greater than that of the detected line (with an average
width of 15$\pm$1.2\AA).  Similar results are found when testing a few
other plausible ratios of the two [OII] components.  We can therefore
rule out that the line is [OII]$\lambda\lambda$372.6,372.9 at z$=$2.12
at a level greater than 99.9\% or 3-$\sigma$ because it produces a line
that is clearly broader than that measured.

\begin{figure}
\centerline{\includegraphics[width=5in]{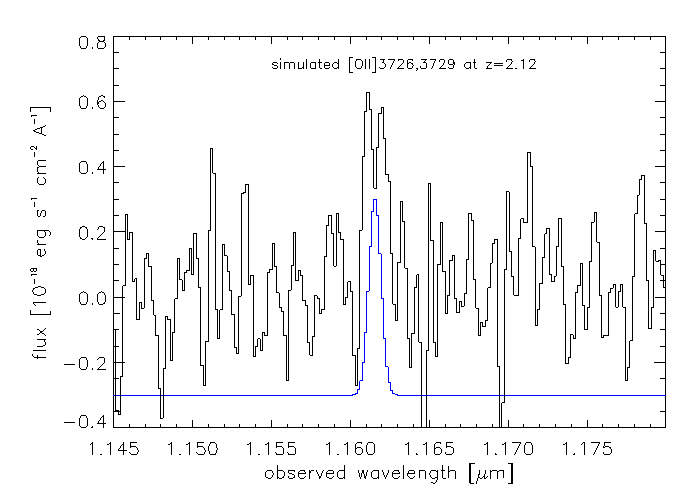}}
\caption{An alternative identification for the line as arising
from the doublet of singly ionised Oxygen. To illustrate that
it is unlikely that the identified line is consistent with being
[OII]$\lambda\lambda$372.6,372.9 at lower redshift we subtracted our
best fit Gaussian profile for the data cube to remove the emission
from the source.  This produces a line-free spectrum with the noise
characteristics for our data cube in the region of and near the
wavelength of the line.  At line centre, we added two delta functions
separated by 2.8(1+z$_{[OII]}$)\AA\ which have been convolved with a
Gaussian profile.  This profile has a width that is exactly the same as
the spectral resolution of our data.  For this illustration we assumed
a line ratio R$_{3726/3729}=$1 and a redshift of 2.12. This redshift
corresponds approximately to the secondary minimum in the redshift
probability distribution derived from the WFC3 photometry$^{40}$. The
blue line shows the Gaussian fit to our data shifted along the ordinate
for comparison. Clearly the doublet would be resolved spectrally and
overall be broader than the detected emission line even at the low
signal-to-noise of our detection. }
\label{fig:o2}
\end{figure}

From the photometry and the line flux, the observed equivalent width
of the line is about 1900\AA. All hypotheses other than Ly$\alpha$
at z$=$8.55 also imply implausibly high equivalent widths for optical
lines of $\approx$600 to 1,100\AA, which are not observed in the
integrated spectra of local galaxies$^{41}$ or even the majority of
the rare photometrically selected galaxies that have extremely high
equivalent widths$^{42}$.  Even if we include a systematic uncertainty
in the flux of several tens of percent, up to 40\%, this would still be
highly implausible. We therefore conclude that the line is indeed most
likely Ly-$\alpha$ at z$=$8.55.

We finally conclude that the line detected is not spurious, is
inconsistent with an instrumental artefact, and is inconsistent with
being an emission line at lower redshift.


\begin{thebibliography}{10}

\bibitem[1]{} Mesinger, A., \& Furlanetto, S., Efficient Simulations
of Early Structure Formation and Reionization, {\it Astrophysical
Journal}, {\bf 669}, 663-675, (2007).

\bibitem[2]{choudhury09}Choudhury, T. R., Haehnelt, M. G., \& Regan,
J., Inside-out or outside-in: the topology of reionization in the
photon-starved regime suggested by Ly-alpha forest data, {\it Monthly
Notices of the Royal Astronomical Society}, {\bf 394}, 960-977, (2009).

\bibitem[3]{komatsu10} Komatsu, E. et al., Seven-Year
Wilkinson Microwave Anisotropy Probe (WMAP) Observations:
Cosmological Interpretation, {\it Astrophys. J. Suppl., submitted
(astro-ph/1001.4538)}, (2010)

\bibitem[4]{fan06} Fan, X., Strauss, M. A., Becker, R. H., White,
R. L., Gunn, J. E., Knapp, G. R., Richards, G. T., Schneider, D. P.,
Brinkmann, J., \& Fukugita, M., Constraining the Evolution of the Ionizing
Background and the Epoch of Reionization with z~6 Quasars. II. A Sample
of 19 Quasars, {\it Astronomical Journal}, {\bf 132}, 117-136, (2006).

%5
\bibitem[5]{bouwens10} Bouwens, R. J., Illingworth, G. D., Oesch, P. A.,
Stiavelli, M., van Dokkum, P., Trenti, M., Magee, D., Labb\'{e}, I.,
Franx, M., Carollo, C. M., \& Gonzalez, V., Discovery of z$\sim$8 Galaxies
in the Hubble Ultra Deep Field from Ultra-Deep WFC3/IR Observations,
{\it Astrophysical Journal}, {\bf 709}, L133-L137, (2010).

\bibitem[6]{} Tanvir, N. R., et al., A $\gamma$-ray burst at a redshift
of z$\sim$8.2, {\it Nature}, {\bf 461}, 1254-1257, (2009).

\bibitem[7]{} Salvaterra, R., et al., GRB090423 at a
redshift of z$\sim$8.1, {\it Nature}, {\bf 461}, 1258-1260, (2009).

\bibitem[8]{} Iye, M., Ota, K., Kashikawa, N., Furusawa, H., Hashimoto,
T., Hattori, T., Matsuda, Y., Morokuma, T., Ouchi, M., \& Shimasaku, K.,
A galaxy at a redshift z = 6.96, {\it Nature}, {\bf 443}, 186-188, (2006).

\bibitem[9]{loeb05} Loeb, A., Barkana, R., \& Hernquist, L., Was the
Universe Reionized at Redshift 10?,{\it Astrophysical Journal}, {\bf 620},
553-558, (2005).

\bibitem[10]{mclure10} McLure, R. J., Dunlop, J. S., Cirasuolo, M.,
Koekemoer, A. M., Sabbi, E., Stark, D. P., Targett, T. A., \& Ellis,
R. S.,  Galaxies at z = 6-9 from the WFC3/IR imaging of the Hubble Ultra
Deep Field,  {\it Monthly Notices of the Royal Astronomical Society},
 {\bf 403}, 960-983,  (2010).

\bibitem[11]{ouchi08} Ouchi, M., Shimasaku, K., Akiyama, M., Simpson,
C., Saito, T., Ueda, Y., Furusawa, H., Sekiguchi, K., Yamada, T., Kodama,
T., Kashikawa, N., Okamura, S., Iye, M., Takata, T., Yoshida, M., \&
Yoshida, M.,  The Subaru/XMM-Newton Deep Survey (SXDS). IV. Evolution of
Ly$\alpha$ Emitters from z=3.1 to 5.7 in the 1 deg$^2$ Field: Luminosity
Functions and AGN,  {\it Astrophysical Journal Supplement Series},
{\bf 176}, 301-330,  (2008).

\bibitem[12]{ota08} Ota, K., Iye, M., Kashikawa, N., Shimasaku, K.,
Kobayashi, M., Totani, T., Nagashima, M., Morokuma, T., Furusawa, H.,
Hattori, T., Matsuda, Y., Hashimoto, T., \& Ouchi, M.,  Reionization and
Galaxy Evolution Probed by z = 7 Ly$\alpha$ Emitters,  {\it Astrophysical
Journal}, {\bf 677}, 12-26,  (2008).

\bibitem[13]{ouchi09} Ouchi, M., Mobasher, B., Shimasaku, K., Ferguson,
H. C., Fall, S. M., Ono, Y., Kashikawa, N., Morokuma, T., Nakajima, K.,
Okamura, S., Dickinson, M., Giavalisco, M., \& Ohta, K.,  Large Area
Survey for z = 7 Galaxies in SDF and GOODS-N: Implications for Galaxy
Formation and Cosmic Reionization,  {\it Astrophysical Journal}, {\bf
706}, 1136-1151,  (2009).

\bibitem[14]{castellano10} Castellano, M., et al.,  Evidence of a fast
evolution of the UV luminosity function beyond redshift 6 from a deep
HAWK-I survey of the GOODS-S field, {\it Astronomy and Astrophysics},
{\bf 511}, A20-36, (2010).

\bibitem[15]{verma07} Verma, A., Lehnert, M. D., F{\"o}rster Schreiber,
N. M., Bremer, M. N., \& Douglas, L., Lyman-break galaxies at z$\sim$5 -
I. First significant stellar mass assembly in galaxies that are not
simply z$\sim$3 LBGs at higher redshift, {\it Monthly Notices of the Royal
Astronomical Society}, {\bf 377}, 1024-1042,  (2007).

\bibitem[16]{stark09} Stark, D. P., Ellis, R. S., Bunker, A., Bundy,
K., Targett, T., Benson, A., \& Lacy, M.,  The Evolutionary History of
Lyman Break Galaxies Between Redshift 4 and 6: Observing Successive
Generations of Massive Galaxies in Formation,  {\it Astrophysical
Journal}, {\bf 697}, 1493-1511, (2009).

%20
\bibitem[17]{bouwens10b} Bouwens, R. J., Illingworth, G. D., Oesch,
P. A., Trenti, M., Stiavelli, M., Carollo, C. M., Franx, M., van Dokkum,
P. G., Labb{\'e}, I., \& Magee, D.,  Very Blue UV-Continuum Slope $\beta$
of Low Luminosity z$\sim$7 Galaxies from WFC3/IR: Evidence for Extremely Low
Metallicities?,{\it Astrophysical Journal}, {\bf 708}, L69-L73, (2010).

\bibitem[18]{} Bromm, V., Kudritzki, R. P., \& Loeb, A., Generic
Spectrum and Ionization Efficiency of a Heavy Initial Mass Function
for the First Stars, {\it Astrophysical Journal}, {\bf 552},464-472,
(2001).

\bibitem[19]{} Labb\'{e}, I., Gonz\'{a}lez, V., Bouwens, R. J.,
Illingworth, G. D., Franx, M., Trenti, M., Oesch, P. A., van Dokkum,
P. G., Stiavelli, M., Carollo, C. M., Kriek, M., \& Magee, D., Star
Formation Rates and Stellar Masses of z = 7-8 Galaxies from IRAC
Observations of the WFC3/IR Early Release Science and the HUDF Fields,
{\it Astrophysical Journal}, {\bf 716}, L103-L108, (2010).

\bibitem[20]{furlanetto08} Furlanetto, S. R., Haiman, Z., \& Oh,
S. P.,  Fossil Ionized Bubbles around Dead Quasars during Reionization,
{\it Astrophysical Journal}, {\bf 686}, 25-40,  (2008).

\bibitem[21]{miralda-escude98}Miralda-Escude, J., Reionization of the
Intergalactic Medium and the Damping Wing of the Gunn-Peterson Trough,
{\it Astrophysical Journal}, {\bf 501}, 15-22, (1998).

\bibitem[22]{wyithe05} Wyithe, J. S. B., \& Loeb, A.,  Undetected
Sources Allow Transmission of the Ly$\alpha$ Line from Galaxies Prior to
Reionization,  {\it Astrophysical Journal}, {\bf 625}, 1-5, (2005).

\bibitem[23]{}Dijkstra, M., \& Wyithe, S., Seeing Through the Trough:
Outflows and the Detectability of Lyman Alpha Emission from the First
Galaxies, (astro-ph/1004.2490) submitted to Monthly Notices of the Royal
Astronomical Society, (2010).

\bibitem[24]{wyithe04} Wyithe, J. S. B., \& Loeb, A.,  A characteristic
size of $\sim$10 Mpc for the ionized bubbles at the end of cosmic
reionization, {\it Nature}, {\bf 432}, 194-196,  (2004).

\bibitem[25]{} Miralda-Escud{\'e}, J., Haehnelt, M., \& Rees,
M. J., Reionization of the Inhomogeneous Universe, {\it Astrophysical
Journal}, {\bf 530}, 1-16, (2000).

\bibitem[26]{}Scalo, J., The IMF Revisited: A Case for Variations,
{\it The Stellar Initial Mass Function (38th Herstmonceux Conference)},
ASP conference series 142,201-209, (1998). ed. G. Gilmore and D. Howell
(San Francisco: ASP)

\bibitem[27]{deharveng01}Deharveng, J.-M., Buat, V., Le Brun, V.,
Milliard, B., Kunth, D., Shull, J.  M., \& Gry, C.,  Constraints on the
Lyman continuum radiation from galaxies: First results with FUSE on Mrk
54,{\it Astronomy and Astrophysics}, {\bf 375}, 805-813, (2001).

\bibitem[28]{inoue05} Inoue, A. K., Iwata, I., Deharveng, J.-M., Buat,
V., \& Burgarella, D.,  VLT narrow-band photometry in the Lyman continuum
of two galaxies at z$\sim$3. Limits to the escape of ionizing flux,
{\it Astronomy and Astrophysics}, {\bf 435}, 471-482, (2005).

\bibitem[29]{bergvall06} Bergvall, N., Zackrisson, E., Andersson,
B.-G., Arnberg, D., Masegosa, J., \& \"{O}stlin, G., First detection of
Lyman continuum escape from a local starburst galaxy. I. Observations
of the luminous blue compact galaxy Haro 11 with the Far Ultraviolet
Spectroscopic Explorer (FUSE), {\it Astronomy and Astrophysics}, {\bf 448},
513-524, (2006).

\end{thebibliography}

\begin{thebibliography}{10}

\bibitem[31]{} Bonnet, H., et al.,  First light of SINFONI at
the VLT,  {\it The Messenger}, {\bf 117}, 17-24,  (2004).

\bibitem[32]{} Bouwens, R. J., Illingworth, G. D., Oesch, P. A.,
Stiavelli, M., van Dokkum, P., Trenti, M., Magee, D., Labb\'{e}, I.,
Franx, M., Carollo, C. M., \& Gonzalez, V., Discovery of z$\sim$8 Galaxies
in the Hubble Ultra Deep Field from Ultra-Deep WFC3/IR Observations,
{\it Astrophysical Journal},{\bf 709}, L133-L137, (2010).

\bibitem[33]{} Tody, D. 1993, in Astronomical Data Analysis
Software and Systems II, ed.  R. J. Hanisch, R. J. V. Brissenden, \&
J. Barnes (San Francisco: ASP), 173

\bibitem[34]{} Nesvadba, N. P. H., Lehnert, M. D., De Breuck,
C., Gilbert, A. M., \& van Breugel, W.,  Evidence for powerful AGN winds
at high redshift: dynamics of galactic outflows in radio galaxies during
the ``Quasar Era'',  {\it Astronomy and Astrophysics}, {\bf 491}, 407-424,
(2008).

\bibitem[35]{} Nesvadba, N. P. H., Lehnert, M. D., Davies,
R. I., Verma, A., \& Eisenhauer, F.,  Integral-field spectroscopy of a
Lyman-break galaxy at z = 3.2: evidence for merging,{\it Astronomy and
Astrophysics},{\bf 479}, 67-73, (2008).

\bibitem[36]{} Nesvadba, N. P. H., Lehnert, M. D., Genzel, R.,
Eisenhauer, F., Baker, A.  J., Seitz, S., Davies, R., Lutz, D., Tacconi,
L., Tecza, M., Bender, R., \& Abuter, R.,  Intense Star Formation and
Feedback at High Redshift: Spatially Resolved Properties of the z = 2.6
Submillimeter Galaxy SMM J14011+0252,  {\it Astrophysical Journal},
{\bf 657}, 725-737,  (2007).

\bibitem[37]{} Lehnert, M. D., Nesvadba, N. P. H., Tiran,
L. L., Matteo, P. D., van Driel, W., Douglas, L. S., Chemin, L.,
\& Bournaud, F.,  Physical Conditions in the Interstellar Medium of
Intensely Star-Forming Galaxies at Redshift~2,  {\it Astrophysical
Journal},{\bf 699}, 1660-1678, (2009).

\bibitem[38]{} Richard, J., Stark, D. P., Ellis, R. S., George, M. R.,
Egami, E., Kneib, J.-P., \& Smith, G. P., A Hubble and Spitzer Space
Telescope Survey for Gravitationally Lensed Galaxies: Further Evidence
for a Significant Population of Low-Luminosity Galaxies beyond z=7,
{\it Astrophysical Journal},{\bf 685}, 705-724, (2008).

\bibitem[39]{} Douglas, L. S., Bremer, M. N., Lehnert, M. D., Stanway,
E. R., \& Milvang-Jensen, B., Spectroscopy of z$\sim$5 Lyman Break
Galaxies in the ESO Remote Galaxy Survey, {\it Monthly Notices of the
Royal Astronomical Society}, accepted (2010).

\bibitem[40]{}  McLure, R. J., Dunlop, J. S., Cirasuolo, M.,
Koekemoer, A. M., Sabbi, E., Stark, D. P., Targett, T. A., \& Ellis,
R. S.,  Galaxies at z = 6-9 from the WFC3/IR imaging of the Hubble Ultra
Deep Field,  {\it Monthly Notices of the Royal Astronomical Society},
{\bf 403}, 960-983,  (2010).

\bibitem[41]{} Kennicutt, R. C., Jr.,  The integrated
spectra of nearby galaxies - General properties and emission-line spectra,
{\it Astrophysical Journal}, {\bf 388}, 310-327,  (1992).

\bibitem[42]{} Cardamone, C., Schawinski, K., Sarzi, M., Bamford, S. P.,
Bennert, N., Urry, C. M., Lintott, C., Keel, W. C., Parejko, J., Nichol,
R. C., Thomas, D., Andreescu, D., Murray, P., Raddick, M. J., Slosar,
A., Szalay, A., \& Vandenberg, J., Galaxy Zoo Green Peas: discovery of
a class of compact extremely star-forming galaxies, {\it Monthly Notices
of the Royal Astronomical Society}, {\bf 399}, 1191-1205, (2009).

\end{thebibliography}
\end{document}